\documentclass[fleqn,usenatbib]{mnras}
\usepackage{newtxtext,newtxmath}
\usepackage[T1]{fontenc}
\usepackage{graphicx}	% Including figure files
\usepackage{amsmath}	% Advanced maths commands
\pubyear{2023}
\usepackage[utf8]{inputenc}

\usepackage{booktabs}
\usepackage{tabularx}
\usepackage{subcaption}
\usepackage[capitalise]{cleveref}
\usepackage{siunitx}
\usepackage{orcidlink}

\newcommand{\vecA}{\vec{\mathcal{A}}}
\newcommand{\vecl}{\vec{\mathcal{\lambda}}}
\newcommand{\der}[2]{\frac{d#2}{d#1}}
\newcommand{\PP}{$P$--$\dot{P}$}

\title[Populations, parameters of NSs with CWs and 3G detectors]{%
Population synthesis and parameter estimation of neutron stars with continuous gravitational waves and third-generation detectors%
}
\author[%
Yuhan Hua, Karl Wette, Susan M. Scott, Matthew D. Pitkin%
]{%
Yuhan Hua,$^{\orcidlink{0009-0007-1449-2383}\,1,2}$\thanks{E-mail: u6982969@alumni.anu.edu.au} %
Karl Wette,$^{\orcidlink{0000-0002-4394-7179}\,1,2}$\thanks{E-mail: karl.wette@anu.edu.au}, %
Susan M. Scott$^{\orcidlink{0000-0002-9875-7700}\,1,2}$, %
and %
Matthew D. Pitkin$^{\orcidlink{0000-0003-4548-526X}\,3,4}$
\\
$^1$Centre for Gravitational Astrophysics, Australian National University, Canberra ACT 2601, Australia\\
$^2$ARC Centre of Excellence for Gravitational Wave Discovery (OzGrav), Hawthorn VIC 3122, Australia\\
$^3$CEDAR Audio Ltd, Cambridge, CB21 5BS, United Kingdom\\
$^4$SUPA, University of Glasgow, Glasgow G12 8QQ, United Kingdom
}
\date{\today}

\begin{document}
\label{firstpage}
\pagerange{\pageref{firstpage}--\pageref{lastpage}}
\maketitle

\begin{abstract}
    Precise measurement of stellar properties through the observation of continuous gravitational waves from spinning non-axisymmetric neutron stars can shed light onto new physics beyond terrestrial laboratories. Although hitherto undetected, prospects for detecting continuous gravitational waves improve with longer observation periods and more sensitive gravitational wave detectors. We study the capability of the Advanced Laser Interferometer Gravitational-Wave Observatory, and the Einstein Telescope to measure the physical properties of neutron stars through continuous gravitational wave observations. We simulate a population of Galactic neutron stars, assume continuous gravitational waves from the stars have been detected, and perform parameter estimation of the detected signals. Using the estimated parameters, we infer the stars' moments of inertia, ellipticities, and the components of the magnetic dipole moment perpendicular to the rotation axis. The estimation of the braking index proved challenging and is responsible for the majority of the uncertainties in the inferred parameters. Using the Einstein Telescope with an observation period of $\SI{5}{yrs}$, point estimates using median can be made on the moments of inertia with error of $\sim 10-100\%$ and on the ellipticities with error of $\sim 5-50\%$, subject to the inference of the braking index. The perpendicular magnetic dipole moment could not be accurately inferred for neutron stars that emit mainly gravitational waves.
\end{abstract}

\begin{keywords}
    stars: neutron - gravitational waves - pulsars: general
\end{keywords}

\section{Introduction}
One hundred years after \citet{einsteinFeldgleichungenGravitation1915} published his paper on the general theory of relativity, the first gravitational wave detection \citep[the binary black hole merger GW150914,][]{abbottObservationGravitationalWaves2016} was made on September 14, 2015 by the Advanced Laser Interferometer Gravitational-Wave Observatory \citep[LIGO,][]{aasiAdvancedLIGO2015}, a major milestone for gravitational wave astronomy. 
The observation of the binary neutron star merger GW170817 by LIGO and Virgo \citep{acerneseAdvancedVirgoSecondgeneration2014} in both gravitational and electromagnetic radiation \citep{abbottGW170817ObservationGravitational2017,abbottGravitationalWavesGammaRays2017} is an example of multi-messenger astronomy and offers hope to further our understanding of neutron stars, which are known for their extreme densities and currently poorly understood physics.

Gravitational waves from mergers of binary neutron stars only convey information at the end of their life cycle, when they are under strong mutual gravitational perturbation. Alternatively, a neutron star with asymmetries about its axis of rotation can generate continuous gravitational waves (henceforth ``continuous waves'') that are long-lived and quasi-monochromatic \citep{zimmermannGravitationalWavesRotating1979,bonazzolaGravitationalWavesPulsars1996,rilesRecentSearchesContinuous2017,sieniawskaContinuousGravitationalWaves2019}.
As continuous waves are generated by neutron stars in their equilibrium state, they could provide complementary insights into the stars’ physical properties not conveyed through binary neutron star mergers. 
Continuous waves are weaker than gravitational waves from mergers, however, and have yet to be detected by the current second generation detectors LIGO and Virgo \citep{Piccinni_2022,Riles_2023}.

Efforts to detect continuous waves from neutron stars include development of, and improvements to, data analysis techniques for various types of searches \citep{tenorioSearchMethodsContinuous2021,wetteSearchesContinuousGravitational2023}. These are typically specialised to a variety of sources, including but not limited to targeted searches and narrow-band searches for known pulsars \citep{abbottNarrowbandSearchesContinuous2022,Zhang_2021}, directed searches for supernova remnants \citep{abbottSearchesContinuousGravitational2021,Lindblom_2020}, and all-sky surveys for undiscovered neutron stars \citep{abbottAllskySearchContinuous2019,abbottAllskySearchContinuous2022,Covas_2022}. 
Directly increasing the sensitivities of gravitational wave detectors by further suppressing the noise sources also improves the prospect of a first detection.
Current upgrades to the second generation detectors are anticipated to reduce the noise amplitude spectral density by a factor of two \citep{millerProspectsDoublingRange2015}.

The planned third generation detectors are expected to achieve a tenfold increase in strain sensitivity in a wide frequency range, offering hope to the detection of gravitational waves from other sources beyond binary compact object mergers \citep{sathyaprakashScientificObjectivesEinstein2012,bailesGravitationalwavePhysicsAstronomy2021,hallCosmicExplorerNextGeneration2022}.
Proposed third generation ground-based detector concepts include the Einstein Telescope (ET) and the Cosmic Explorer, both of which are planned to begin construction in the 2030s. The Einstein Telescope is expected to have three arms of 10 km length, arranged in an equilateral triangle formation and located underground. It will have three detectors at the vertices of the triangle and two interferometers at each detector, forming a total of six interferometers \citep{Punturo_2010,sathyaprakashScientificObjectivesEinstein2012}. 
Cosmic Explorer is planned to be constructed on the surface with a design similar to that of LIGO, but with arm length increased tenfold to \SI{40}{km} \citep{reitzeCosmicExplorerContribution2019,evans2021horizon}.

In this work we present a study of the potential for observations of continuous waves from isolated neutron stars by LIGO and the third generation detector Einstein Telescope, and the estimation of the continuous wave signal parameters. 
In addition, we also estimate the errors from the inference of neutron star properties, including the principal moment of inertia, ellipticity, and the component of magnetic dipole moment perpendicular to the rotation axis, using a theoretical framework developed by \citet{luInferringNeutronStar2023}.
After reviewing relevant background in \cref{sec:background}, we synthesise a population of neutron stars emitting continuous waves and electromagnetic radiation in \cref{sec:synthesis}, and infer the stellar properties of ten neutron stars with the largest continuous wave characteristic strain amplitude using Bayesian inference in \cref{sec:param-est}. We conclude with a discussion in \cref{sec:discussion}.

\section{Background}\label{sec:background}
Currently, neutron stars are observed primarily as pulsars. A pulsar emits electromagnetic radiation, typically directed along its magnetic dipole axis as a result of a strong stellar magnetic field ($\sim \SI{e 12}{G}$). The open field lines of the magnetic field are strong enough to accelerate charges to a relativistic speed sufficient to escape the magnetosphere  \citep{kramerPulsars2005}. 
If the dipole axis of the magnetic field is not aligned with the axis of rotation, the accelerated charges produce regular pulses of electromagnetic radiation at the same frequency as the rotation of the pulsar. A pulsar may be thought of as an extraterrestrial “lighthouse” that emits radiation at a fixed location for an observer. 
Pulsar evolution is well described using its period $P$ and spindown $\dot{P}$, as both its characteristic age
\begin{align}\label{eqn:charAge}
    t_{\text{char.}}=\frac{P}{(n-1) \dot{P}}
\end{align}
and magnetic field strength
\begin{align}\label{eqn:charB}
    B=\SI{3.2e 19}{} \sqrt{P\dot{P}} ~\si{G}
\end{align}
depend only on these two parameters \citep{kramerPulsars2005}. (The braking index $n$ is defined and explained below in \cref{eqn:braking}.)

In addition to electromagnetic radiation, neutron stars may also emit continuous waves through several mechanisms, such as deformation away from axi-symmetry due to their magnetic fields \citep{zimmermannGravitationalWavesRotating1979,bonazzolaGravitationalWavesPulsars1996}, quasi-normal fluid perturbations known as $r$-modes \citep{owenGravitationalWavesHot1998,1998ApJ...502..708A,1998ApJ...502..714F}, or accretion of matter from a binary companion star \citep{Bild1998-GrvRdtRtAcNtS,WattEtAl2008-DtcGrvWEmKAcNtS}. Deformation of neutron stars about their rotation axes can be caused by cooling and cracking of the crust \citep{pandharipandeNeutronStarStructure1976}, a non-axisymmetric magnetic field \citep{zimmermannGravitationalWavesRotating1979}, magnetically-confined mountains \citep{melatosGravitationalRadiationAccreting2005}, or electron capture gradients \citep{ushomirskyDeformationsAccretingNeutron2000}.
The characteristic strain amplitude of a continuous wave signal from a deformed neutron star is given by \citep{jaranowskiDataAnalysisGravitationalwave1998}
\begin{align}\label{eqn:h0}
    h_0=\frac{4\pi^2G}{c^4} \frac{I_{zz}\epsilon}{r}f^2 \,,
\end{align}
where $G$ is the gravitational constant; $c$ is the speed of light; $I_{zz}$ is the principal moment of inertia aligned with the rotation axis; $\epsilon$ is the equatorial ellipticity that characterises the extent of the neutron star's deformation; $r$ is the distance to the detector; and $f=2\nu$ is the gravitational wave frequency, taken to be twice the rotational frequency $\nu$. 

As a neutron star rotates, it may emit energy through electromagnetic and gravitational wave radiation, which extracts rotational kinetic energy and in turn reduces its spin frequency. The spindown of a neutron star can be characterised as \citep{manchesterSecondMeasurementPulsar1985}
\begin{align}\label{eqn:spindown}
    \dot{\nu}=-K\nu^n,
\end{align}
where $K$ is a constant, and the braking index $n$ is obtained by differentiating and rearranging \cref{eqn:spindown}:
\begin{align}\label{eqn:braking}
    n=\frac{\nu \ddot{\nu}}{\dot{\nu}^2}.
\end{align}
It is expected from theory that neutron stars have a braking index ranging between 3 and 5, with 3 being the case when the neutron star emits only electromagnetic radiation \citep{ostrikerNaturePulsarsTheory1969}, and 5 being the case for only continuous wave emission through time-varying mass quadrupole. Gravitational waves can additionally be emitted through a current-type quadrupole moment due to r-modes, which results in a breaking index of 7. 
There are measured braking indices from pulsars that span orders of magnitude outside this range \citep{johnstonPulsarBrakingIndices1999,zhangWHYBRAKINGINDICES2012,lowerImpactGlitchesYoung2021}, however this likely reflects the difficulty of accurately measuring the second derivative $\ddot{\nu}$ of the rotational frequency, or the impact of glitches \citep{10.1093/mnras/staa2640} or timing noise \citep{Vargas_2023,10.1111/j.1365-2966.2004.08157.x}. 
We adopt the commonly accepted range of $3\le n\le 5$. 

The amplitude of a continuous wave signal observed by a detector from a neutron star will be modulated by the detector's antenna pattern. We henceforth assume the neutron stars to be biaxial rotors with continuous wave frequency of $f=2\nu$ and amplitude given by
\citep{jaranowskiDataAnalysisGravitationalwave1998}:
\begin{align}\label{eqn:ht}
    h(t)&=h_0\bigg\{ \frac{1}{2}(1+\cos^2\iota)F_+(t,\psi)\cos[\phi_0+\phi(t)]\nonumber\\
        &\quad+\cos\iota F_\times (t,\psi) \sin[\phi_0+\phi(t)] \bigg\},
\end{align}
where $h_0$ is the characteristic strain amplitude; $\phi(t)$ is the phase of the signal with initial value $\phi_0$; and $\cos\iota$ is the orientation angle, describing whether the neutron star is viewed face-on (spin-axis along the line of sight, $\cos\iota=\pm 1$) or edge-on (spin-axis perpendicular to the line of sight, $\cos\iota=0$). The $F_{+,\times }$ are the detector response functions for $+$ or $\times$ polarised gravitational waves respectively, and are parameterised by the polarisation angle $\psi$.  The parameters $\vecA\equiv (h_0,\cos\iota,\psi,\phi_0)$ are commonly referred to as the amplitude parameters.

The continuous wave signal of an isolated neutron star is expected to be nearly monochromatic in the neutron star reference frame; with respect to the detector reference frame, however, it is Doppler shifted by the rotation and the orbital motion of the Earth. 
On the time-scale of a day, the signal is modulated by the rotation of the Earth; while on the time-scale of a year, the signal is modulated by the orbit of the Earth around the Solar System Barycentre (SSB).
(The velocity of the SSB relative to the neutron star could further affect the detected frequency, although this effect is approximately constant on the timescale of measurements.) 
In the neutron star frame of reference, the continuous wave signal phase is modelled, up to the second-order time derivative, as \citep{jaranowskiDataAnalysisGravitationalwave1998}
\begin{align}\label{eqn:phi}
    \phi(\tau)=2\pi \left[f\tau+\frac{1}{2}\dot{f}\tau^2+\frac{1}{6}\ddot{f}\tau^3+\mathcal{O}(\tau^4)\right],
\end{align}
where $\tau$ is time in the SSB frame; and $f,\dot{f},\ddot{f}$ are the continuous wave frequency and its first- and second-order time derivatives, or spindowns, at a reference time $\tau=0$.
The neutron star time $\tau$ can be converted to time $t$ in the detector's frame of reference by
\begin{align}
    \tau(t)=t+\frac{\vec{r}\cdot \hat{n}}{c},
\end{align}
where $\vec{r}$ is the position of the detector with respect to the SSB, and $\hat{n}$ is the unit vector pointing from the SSB to the neutron star. (Here we ignore relativistic effects, and assume the neutron star is at rest with respect to the SSB.) The parameters $\vecl\equiv \left( f,\dot{f},\ddot{f},\hat{n} \right)$ are typically referred to as the phase evolution parameters.

\section{Neutron star population synthesis}\label{sec:synthesis}

Since no continuous waves have been detected from the Galactic neutron star population, a simulated neutron star population must realistically contain stars that emit electromagnetic and/or continuous gravitational waves.
Such a population has been previously investigated by a number of authors \citep{palombaSimulationPopulationIsolated2005,knispelBlandfordArgumentStrongest2008,wadeContinuousGravitationalWaves2012,pitkinProspectsObservingContinuous2011,woanEvidenceMinimumEllipticity2018,cieslarDetectabilityContinuousGravitational2021,reedModelingGalacticNeutron2021a}. 
In this work we utilise Monte-Carlo methods to simulate a neutron star population. We assign neutron star parameters drawn from theoretical probability distributions, evolve the neutron stars temporally, and use their final spin frequency to simulate continuous wave signals.

\subsection{Simulation method and parameters}\label{sec:sim-method-par}
The initial spin frequency $\nu_0$ can be obtained through the inverse of the initial period $P_0$. 
Three models for the initial period distribution have been proposed \citep{palombaSimulationPopulationIsolated2005,knispelBlandfordArgumentStrongest2008}. The first model is a lognormal distribution with $\bar{P}_0=\SI{5}{ms}$, $\sigma=0.69$,
% \begin{align}\label{eqn:DistP}
%     \mathrm{P}(P_0)=\frac{1}{\sqrt{2\pi}\sigma P_0} \exp \left[ -\frac{1}{2\sigma^2}(\ln P_0-\ln\bar{P}_0^2)\right],
% \end{align}
and excludes all $P_0<\SI{0.5}{ms}$.
The second model uses the same lognormal distribution but with all $P_0<\SI{10}{ms}$ set to $\SI{10}{ms}$; this mimics the potential existence of $r$-modes in young neutron stars, which increases the spin period to $\SI{10}{ms}$\footnote{Reported to be between $\sim5$ and $\sim\SI{15}{ms}$ in \citet{anderssonGravitationalRadiationLimit1999}; taken to be $\SI{10}{ms}$ by \citet{palombaSimulationPopulationIsolated2005}.} once $r$-modes become saturated. The third model simply uses a uniform distribution of $P_0\in [2,15]\,\SI{}{ms}$; this is intended to accommodate both $r$-modes, and the fall-back of matter after the supernova which decreases the period through additional angular momentum \citep{wattsSpinEvolutionNascent2002}. A uniform initial period distribution was favoured by simulations of Galactic binary neutron stars performed in \citet{sgalletta2023binary}. We follow \citet{palombaSimulationPopulationIsolated2005} and \citet{knispelBlandfordArgumentStrongest2008} and adopt the third model for $P_0$.

We assume an average birth rate of one Galactic neutron star every 100 years, as evidenced by the observation of core-collapse supernovae in the Galaxy \citep{diehlRadioactive26AlMassive2006,rozwadowskaRateCoreCollapse2021}. An array of random variables drawn from a Poisson distribution with rate $\lambda=\SI{100}{yrs}$ is generated and cumulatively summed, and the values of the summed array are taken as the ages of the simulated stars. Similar to \citet{palombaSimulationPopulationIsolated2005}, an upper bound of $t_\text{max}=\SI{e 8}{yrs}$ is used, as a compromise between the computer memory required to store a simulated population and realistic expectations of Galactic neutron star ages. 
% As noted by \citet{palombaSimulationPopulationIsolated2005}, since the kick velocities of neutron stars from supernovae are around $\sim \SI{400}{km.s^{-1}}$, the neutron stars will move $\sim \SI{600}{pc}$ with respect to the Sun in \SI{e 6}{yrs}. It follows that neutron stars near the Sun -- those that are most likely to be detectable in continuous waves -- are young. A maximum age of $\SI{e 8}{yrs}$ is therefore conservative.

The magnetic field strength $B$ and ellipticity $\epsilon$ are chosen from log-uniform distributions, in line with \citet{wadeContinuousGravitationalWaves2012}, \citet{knispelBlandfordArgumentStrongest2008}, and \citet{reedModelingGalacticNeutron2021a}. Unlike \citet{cieslarDetectabilityContinuousGravitational2021}, we do not evolve the ellipticity in time since there is no well-evidenced evolution model as yet. The magnetic field $B$ is bounded between $\SI{e8}{G}$ and $\SI{e 15}{G}$ \citep{reiseneggerMagneticFieldsNeutron2001,konarMagneticFieldsNeutron2017}. The range for ellipticity $\epsilon$ is $\num{e-6}$ \citep{horowitzBreakingStrainNeutron2009} to $\num{e-9}$ \citep{woanEvidenceMinimumEllipticity2018}. The moment of inertia $I_{zz}$ is chosen uniformly from the widely accepted range of $\left[ 1,3 \right]\times\SI{e 38}{kg.m^2}$ \citep{molnvikCalculationsMassMoment1985,worleyNuclearConstraintsMoments2008,miaoMomentInertiaPSR2022}.

Instead of evolving neutron stars through the Galactic potential, we populate the Galaxy with stationary neutron stars according to a theoretical spatial distribution, similar to the approach used by \citet{reedModelingGalacticNeutron2021a}. 
Current understanding suggests that the spatial distribution of Galactic neutron stars follows a Gaussian distribution in the radial direction and a double-sided exponential (Laplace) distribution in the vertical direction \citep{faucher-giguerePulsarContributionGammaray2010,binneyIntroduction2008}. 
In the Galacto-centric cylindrical frame $(\rho,\theta,z)$, the distribution for the radial distance $\rho$ is given by 
\begin{align}\label{eqn:DistRho}
    \mathrm{p}(\rho)= \frac{1}{\sigma  \sqrt{2\pi} }\exp\left( - \frac{\rho^2}{2\sigma^2} \right),
\end{align}
with $\rho>0$ and $\sigma=\SI{5}{kpc}$. The vertical distribution is given by 
\begin{align}\label{eqn:DistZ}
    \mathrm{p}(z)= \frac{1}{2z_0}\exp\left( - \frac{|z|}{z_0} \right),
\end{align}
with $z_0 \in [0.5,1]\,\SI{}{kpc}$ for conventional stars \citep{binneyIntroduction2008}. In their work, \citet{reedModelingGalacticNeutron2021a}, however, chose $z_0=0.1,2,\SI{4}{kpc}$ to investigate the effect of supernova kicks on the distribution of neutron stars. We use $z_0=\SI{2}{kpc}$ as a compromise between a clustered vertical distribution ($z_0=\SI{0.1}{kpc}$) and a spread-out distribution ($z_0=\SI{4}{kpc}$). 
We use a uniform distribution of angle $\theta$ between each neutron star and the Sun with respect to the Galactic centre, with $\theta\equiv0$ for the Sun.
With the location of the Earth assumed to be coincident with the Sun \citep[$\rho_\text{e}=\SI{8.25}{kpc},\theta_\text{e}=0,z_\text{e}=\SI{0.02}{kpc}$;][]{reedModelingGalacticNeutron2021a,humphreysSunDistanceGalactic1995}, the distance $r$ between a neutron star at $( \rho,\theta,z )$ and the Earth can then be calculated. 

One can determine the spindown $\dot{\nu}$ of a neutron star from the conservation of energy, as follows. Assuming the total kinetic energy budget from the star's rotation is expended either in electromagnetic or continuous wave emission gives \citep{luInferringNeutronStar2023}
\begin{align}\label{eqn:EnergyConservation}
    \left( \der{t}{E} \right)_\text{EM}+ \left( \der{t}{E} \right) _\text{GW}=-\left( \der{t}{E} \right)_\text{rot}.
\end{align}
Due to the expected scale of the ellipticity $\epsilon \ll 1$, neutron stars can be assumed to be nearly spherical, and hence the rotational energy can be approximated as that of a rotating sphere \citep{wetteSearchingGravitationalWaves2008}:
\begin{align}\label{eqn:PowerErot}
    \left( \der{t}{E} \right) _\text{rot}&= 4\pi^2 I_{zz} \nu \dot{\nu}.
\end{align}
The energy emitted by a rotating magnetic dipole is \citep{ostrikerNaturePulsarsTheory1969}
\begin{align}\label{eqn:PowerEM}
    \left( \der{t}{E} \right) _\text{EM}&= -\frac{32\pi^4B^2 R^6 \sin^2\alpha \nu^4}{3c^3\mu_0}= -\frac{32\pi^4\mu_0m_p^2 \nu^4}{3c^3},
\end{align}
where $\mu_0$ is the vacuum permeability and $R$ is the radius of the neutron star. The component of the magnetic dipole moment perpendicular to the rotational axis $m_p$ can be related to the magnetic field strength at the surface $B$ by \citep{condonPulsars2016} \footnote{Note that \cref{eqn:m_p} expresses $m_p$ in the standard SI units for a magnetic moment of $\si{A.m^2}$. This equation differs by a factor of $1/\mu_0$ from the equivalent expressions in \citet{condonPulsars2016}, which assume CGS units; and \citet{luInferringNeutronStar2023}, which assume units of $\si{T.m^3}$.}
\begin{align}\label{eqn:m_p}
    m_p=\frac{1}{\mu_0} BR^3 \sin\alpha.
\end{align}
The simulation assumes, for simplicity, that the magnetic dipole moment is entirely perpendicular to the rotation axis, i.e.\ $\sin\alpha=1$.
The power emitted in continuous waves is given by \citep{ostrikerNaturePulsarsTheory1969,rilesRecentSearchesContinuous2017}
\begin{align}\label{eqn:PowerGW}
    \left( \der{t}{E} \right) _\text{GW}&= - \frac{32G}{5c^5}I_{zz}^2 \epsilon^2 \left( 2\pi\nu \right)^6.
\end{align}
Substituting \cref{eqn:PowerEM,eqn:PowerGW} into \cref{eqn:PowerErot} and rearranging yields the equation for the spindown $\dot{\nu}$:
\begin{align}\label{eqn:Nudot}
    \dot{\nu}&= - \frac{512\pi^4 GI_{zz}}{5c^5}\epsilon^2 \nu^5 - \frac{8\pi^2}{3c^3I_{zz}\mu_0} B^2\sin^2\alpha R^6\nu^3.
\end{align}
\citet{wadeContinuousGravitationalWaves2012} solve \cref{eqn:Nudot} and obtain an analytical expression for the neutron star age $t$ as a function of its final and initial spin frequencies, $\nu$ and $\nu_0$ respectively:
\begin{align}\label{eqn:tOfNu}
    t(\nu,\nu_0)= \frac{1}{2|\gamma_\text{EM}|}\left[ \frac{\nu_0^2-\nu^2}{\nu_0^2\nu^2}+\gamma \ln\left( \frac{\nu^2(1+\nu_0^2\gamma)}{\nu_0^2(1+\nu^2\gamma)} \right)  \right],
\end{align}
where
\begin{align}
    \gamma_\text{GW}= - \frac{512\pi^4 GI_{zz}}{5c^5}\epsilon^2,\quad
    \quad \gamma_\text{EM}=- \frac{8\pi^2}{3c^3I_{zz}\mu_0} B^2\sin^2\alpha R^6
\end{align}
and $\gamma=\gamma_\text{GW} / \gamma_\text{EM}$ are used for brevity. 

To obtain the spin frequency evolution $\nu(t)$ of a neutron star, we need to invert \cref{eqn:tOfNu}. 
One approach is to numerically solve \cref{eqn:tOfNu} through root-finding; this method suffers from poor convergence, however, due to the rapid decrease of $\nu$ at small $t$. 
Alternatively, one evaluates \cref{eqn:tOfNu} at an array of sample frequencies $\left\{ \nu_i \right\}$ to obtain the corresponding time array $\left\{ t_i \right\}$, then interpolate to obtain a continuous function of $\nu(t)$. This approach, however, is limited by the range of sample frequencies $\left[ \nu_\text{lower},\nu_\text{upper} \right]$, which corresponds to a range for $\left\{ t_i \right\}$ of $[ t_\text{lower}=t(\nu_\text{upper}),t_\text{upper}=t(\nu_\text{lower}) ]$.
The upper bound for $\left\{ \nu_i \right\}$ can easily be set as $\nu_\text{upper}=\nu_0$, giving a lower bound $t_\text{lower}=t(\nu_\text{upper})=0$. 
The lower bound for $\left\{ \nu_i \right\}$ may, however, be higher than the true value, i.e.\ $t_\text{upper}=t(\nu_\text{lower})<t_\text{age}$, and hence interpolation cannot be performed. 
This occurs for very old stars with large $t_\text{age}$ and very small $\nu$. 
While one can simply lower $\nu_\text{lower}$, doing so requires significantly more points in $\left\{ \nu_i \right\}$ (assuming linear sampling) and is generally inefficient.

\begin{figure}
    \centering
    \includegraphics[width=0.8\columnwidth]{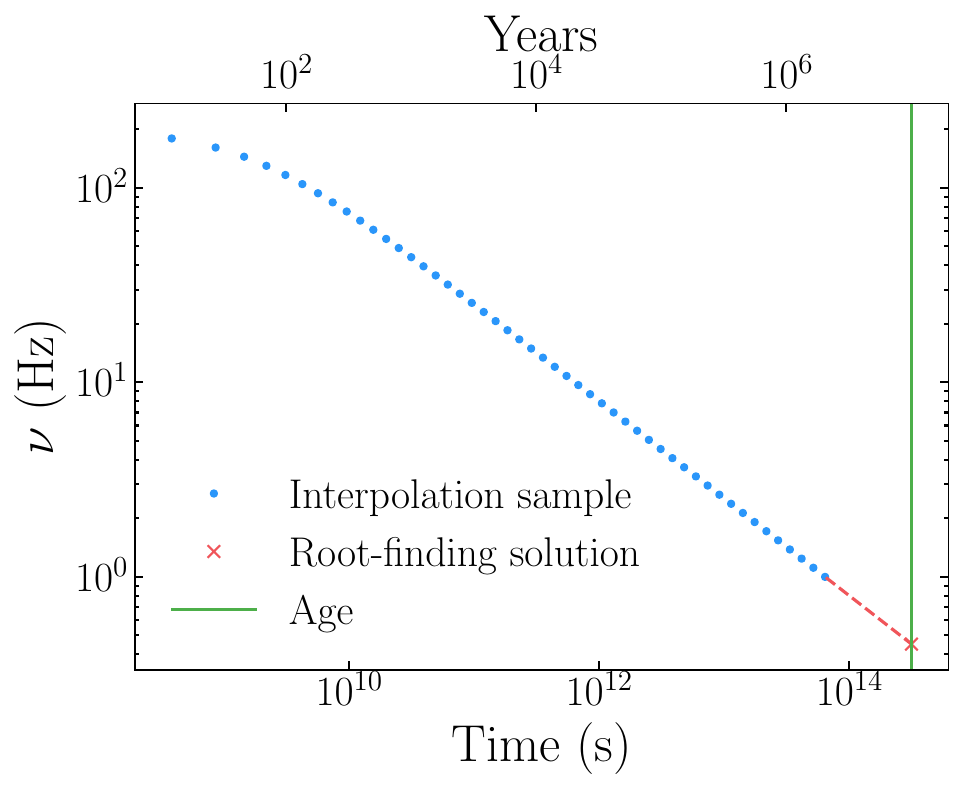}
    \caption{An example of the spin frequency $\nu(t)$ as a function of time. The individual samples from interpolation are plotted in blue. The solution from the root-finding algorithm is marked with a red cross, with the red dashed line showing the extent to which the true age is outside the interpolation samples. The true age is plotted as the green line. }
    \label{fig:nu_evolution}
\end{figure}

We therefore employ a hybrid approach of 
Piece-wise Cubic Hermite Interpolation (PCHIP) at small $t$ and root-finding using the Levenberg-Marquardt algorithm at large $t$ to solve \cref{eqn:tOfNu}. 
\Cref{fig:nu_evolution} illustrates this process. Interpolation is tractable when $\nu(t_\text{age}) \in \left[ \nu_0,0.005\nu_0 \right]$, which is the case for most neutron stars. For older neutron stars with $\nu(t_\text{age}) < 0.005\nu_0$, root-finding is then used, starting with the initial guess $t = t_\text{lower}$, $\nu = 0.005\nu_0$.

\subsection{Synthesised population}
\begin{figure*}
    \begin{subfigure}{\textwidth}
        \centering
        \includegraphics[height=10cm]{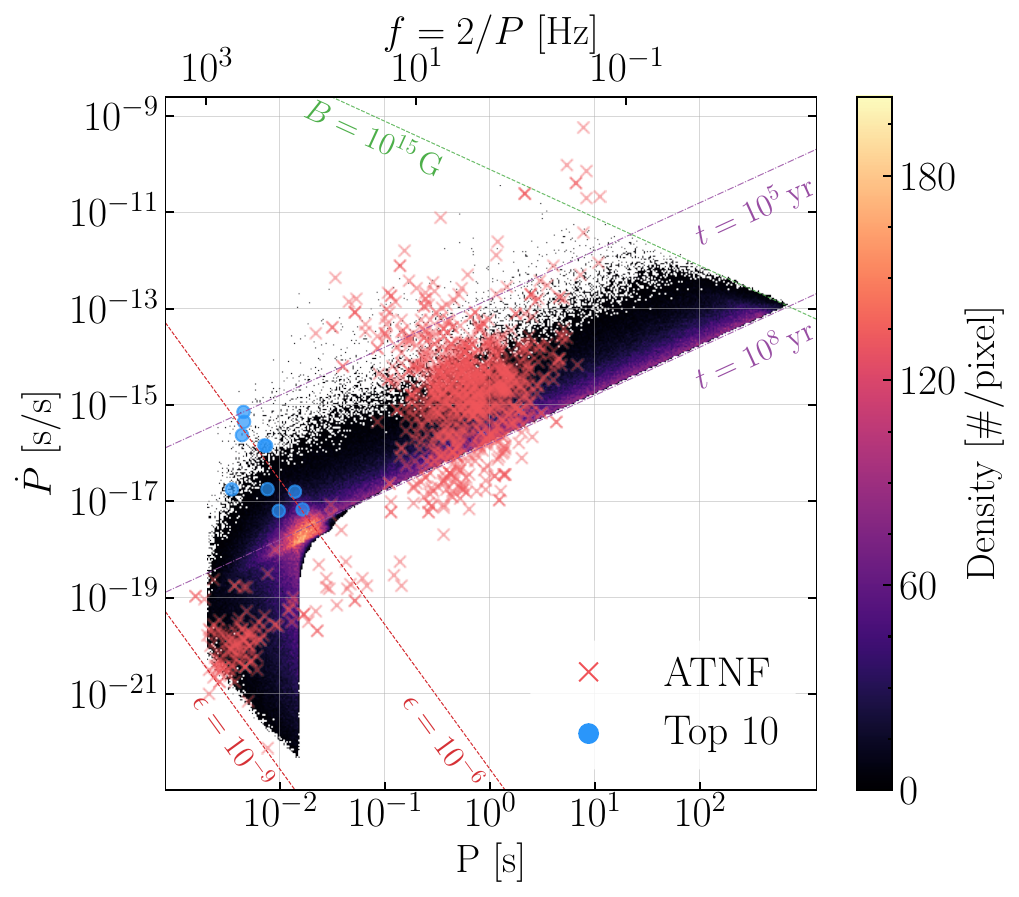}
        \caption{}
        \label{fig:ATNF-PPdot}
    \end{subfigure}\\
    \begin{subfigure}{\textwidth}
        \centering
        \def\svgwidth{11.5cm}
        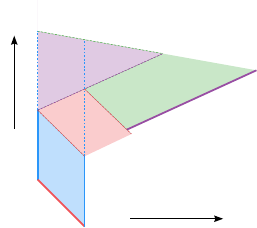
        \caption{}
        \label{fig:PopAnnotation}
    \end{subfigure}
    \caption{(a) \PP\ diagram of a representative synthesised population (coloured pixels) with observational data from the ATNF superimposed (crosses). The top 10 neutron stars with the largest $h_0$ are plotted as blue dots, from which we then perform parameter estimations and will be detailed in \cref{sec:param-est}. A line of constant magnetic field strength $B$ is drawn in green; lines of constant ellipticity $\epsilon$ are drawn in red; lines of constant characteristic age $t$ are drawn in purple. (b) A schematic illustration of four different regions in the \PP\ diagram; see the text for details. Example evolution trajectories from $t = 0$ to $t = t_\text{age}$ for four neutron stars are labelled (1) to (4).}
\end{figure*}

\Cref{fig:ATNF-PPdot} summarises the simulated population, containing \SI{e6}{} neutron stars, in a \PP\ diagram.
The shape of the simulated population is a result of the various assumptions and limitations of the simulation method, as described in \cref{sec:sim-method-par}, and can be understood through \cref{fig:PopAnnotation}.

The population may be partitioned into four regions in the \PP\ space. 
The purple region in \cref{fig:PopAnnotation} contains mainly young neutron stars; they do not have the time to spin down significantly, and hence have higher final spin frequencies (smaller periods) and higher period derivatives than older stars. This positions them in the upper left region of the \PP\ space. Conversely, older stars have more time to spin down, so they have higher periods and lower period derivatives. At a constant birth rate, there are more older neutron stars than young ones, e.g.\ $10$ times more neutron stars with age \SI{e 7}{yrs} than those with age \SI{e 6}{yrs}. This results in a lack of simulated neutron stars in the purple region and a high concentration of older neutron stars.

For a neutron star with negligible ellipticity, its spindown is dominated by electromagnetic radiation, and it follows a trajectory similar to path (1) in \cref{fig:PopAnnotation}. It has a slower spindown of $\dot{\nu}\propto -\nu^3$, corresponding to $\dot{P}\propto P^{-1}$, and will eventually occupy the green region in \cref{fig:PopAnnotation}. 
A neutron star with both significant ellipticity and a strong magnetic field initially radiates energy mostly through continuous waves, then switches to mostly electromagnetic radiation as the spin frequency $\nu$ becomes smaller. Its braking index evolves from $n \sim 5$ to $n \sim 3$, resulting in a non-linear trajectory exemplified by path (2).

The energy emission of a neutron star with significant ellipticity and low magnetic field strength is dominated by continuous wave radiation. Its trajectory is therefore steeper, with spindown $\dot{\nu}\propto -\nu^5$, or $\dot{P}\propto P^{-3}$, as shown by path (3) in \cref{fig:PopAnnotation}. It can reach a slower $\dot{P}$ than stars dominated by electromagnetic radiation with the same characteristic age. This results in the discontinuity in the line of maximum $t_{\text{age}}$ seen in \cref{fig:PopAnnotation}. This can also be seen from \cref{eqn:charAge}, which yields different characteristic ages for stars with different braking indices.  
Given their steeper gradient in the \PP\ plane, these neutron stars are typically located in the red region.

For a neutron star with both insignificant ellipticity and weak magnetic field strength, its spindown will be small, and there will be no significant difference between its initial and final spin frequencies, resulting in a short or even unobservable trajectory demonstrated by path (4) and the stationary star next to it in \cref{fig:PopAnnotation}. The blue region typically occupied by these neutron stars is defined by the minimum and maximum initial period, and the minimum ellipticity.

To be precise, in \cref{fig:PopAnnotation}, the upper bound of the population (green solid and dashed lines) is defined by the maximum magnetic field strength $B_\text{max}$ through \cref{eqn:charB}; the right boundaries of the population (two purple solid lines) are defined by the maximum age through \cref{eqn:charAge} and the maximum ellipticity; the two solid blue lines are defined by the minimum and maximum initial period; and the lower bound of the population (solid red line) is defined by the minimum ellipticity through \cref{eqn:Nudot}.
Inside the population envelope, the red dashed line used to bound the blue region is defined by the intersection of the blue line $P_{0,\text{max}}$ and $t_{\text{max}}$; the intersection of that line with $P_{0,\text{min}}$ is then used to define the purple region; and the red and green regions are then separated by the red dashed line defined by $\epsilon_{\text{max}}$. While these four regions provide a basic intuition regarding the distribution of the population, they do not map one-to-one to all neutron stars. For example, the majority of the continuous wave-dominated (red) region, except near the boundary of $t_\text{max}$, can still contain neutron stars with predominantly electromagnetic radiation. 

The Australian Telescope Network Facility (ATNF) hosts a catalogue of observed pulsars \citep{manchesterAustraliaTelescopeNational2005}, which are superimposed on the synthesised population in \cref{fig:ATNF-PPdot}. 
The majority of the pulsars are clustered at the centre of the diagram, occupying the electromagnetic radiation-dominated (green) region described in \cref{fig:PopAnnotation}.
In addition, the majority of neutron stars are observed at low frequencies $f = 2\nu = 2/P \lesssim \SI{10}{Hz}$, where present-day gravitational wave detectors have limited sensitivity. This is consistent with the current lack of detection of continuous waves through targeted searches of existing pulsars \citep{abbottNarrowbandSearchesContinuous2022,abbottSearchesGravitationalWaves2022}. Note that, while most of the observed pulsars coincide with the simulated population, some pulsars have characteristic ages $> \SI{e 8}{yrs}$. This is a direct consequence of $t_\text{max} = \SI{e 8}{yrs}$ set for the simulation. Regardless, such discrepancies are not likely to drastically affect the usefulness of the simulation in respect of the detectability of continuous waves from young neutron stars.

A subset of pulsars is located in the lower left region of \cref{fig:ATNF-PPdot}, with $P < \SI{0.01}{s}$. These pulsars are millisecond pulsars and are typically in binary systems. Even though they coincide with the low $\dot{P}$ region of the simulated population, the underlying physics of the two are different as we did not consider spin-ups due to accretion in our simulation. However, as argued in \citet{wadeContinuousGravitationalWaves2012}, while we did not explicitly account for millisecond pulsars, we did not exclude them either. An old neutron star that has spun-up through recycling can be thought of as a young neutron star born with a high spin frequency.

Similar to \cref{fig:ATNF-PPdot}, \citet{cieslarDetectabilityContinuousGravitational2021} plotted  simulated neutron stars with continuous waves detectable by the Einstein Telescope in their Fig. 7. Examining the $\dot{P}$ axis, we notice differences between the two sets of results. The detectable neutron stars simulated by \citeauthor{cieslarDetectabilityContinuousGravitational2021} have higher $\dot{P}$, younger characteristic ages and therefore fall in the young (purple) region in \cref{fig:PopAnnotation}. This may be due to their assumption of a decay model for the ellipticity $\epsilon$, which means that for older neutron stars their ellipticities, and consequently $h_0$, are lower. In their model, therefore, only young neutron stars are detectable.

\begin{table*}
\begin{tabularx}{\textwidth}{@{}c@{\extracolsep{\fill}}c@{\extracolsep{\fill}}c@{\extracolsep{\fill}}c@{\extracolsep{\fill}}c@{\extracolsep{\fill}}c@{\extracolsep{\fill}}c@{\extracolsep{\fill}}c@{\extracolsep{\fill}}c@{\extracolsep{\fill}}c@{\extracolsep{\fill}}c@{\extracolsep{\fill}}c@{\extracolsep{\fill}}c@{}c@{\extracolsep{\fill}}}
\toprule
Name & $h_0$ & $\cos\iota$ & $\psi$ & $\phi_0$ & $f$ & $\dot{f}$ & $\ddot{f}$ & $n$ & $I_{zz}$ & $\epsilon$ & $m_p$ & $r$ & Age\\
& & & & & [\si{Hz}] & [\si{Hz/s}] & [\si{Hz/s^2}] & & [\si{kg.m^3}] &  & [\si{Am^2}] & [\si{kpc}] & [\si{Yrs}]\\
\midrule
J1749-0156 & \SI{2.2e-25}{} & -0.56 & 0.71 & 0.77 & \SI{4.4e+02}{} & \SI{-1.4e-12}{} & \SI{2.3e-26}{} & 4.86 & \SI{2.3e+38}{} & \SI{1.5e-07}{} & \SI{9.6e+23}{} & 0.31 & \SI{2.0e+06}{}\\

J1746-0156 & \SI{1.7e-25}{} & -0.69 & 0.40 & 2.20 & \SI{2.1e+02}{} & \SI{-4.0e-14}{} & \SI{3.8e-29}{} & 5.00 & \SI{1.1e+38}{} & \SI{2.2e-07}{} & \SI{4.0e+22}{} & 0.07 & \SI{4.2e+07}{}\\

J1829-0110 & \SI{1.3e-25}{} & 0.98 & 0.32 & 1.74 & \SI{6.4e+02}{} & \SI{-1.3e-10}{} & \SI{1.3e-22}{} & 5.00 & \SI{2.7e+38}{} & \SI{5.0e-07}{} & \SI{1.5e+22}{} & 4.43 & \SI{3.9e+03}{}\\

J1907-0159 & \SI{8.4e-26}{} & 0.61 & 0.10 & 3.02 & \SI{6.3e+02}{} & \SI{-2.3e-10}{} & \SI{4.0e-22}{} & 4.75 & \SI{2.6e+38}{} & \SI{6.8e-07}{} & \SI{9.9e+24}{} & 8.67 & \SI{1.8e+04}{}\\

J1827-0124 & \SI{7.8e-26}{} & -0.92 & 0.79 & 3.01 & \SI{3.9e+02}{} & \SI{-3.7e-11}{} & \SI{1.8e-23}{} & 4.96 & \SI{2.8e+38}{} & \SI{9.2e-07}{} & \SI{3.4e+24}{} & 5.33 & \SI{4.5e+04}{}\\

J1746-0157 & \SI{5.8e-26}{} & -0.95 & 1.45 & 1.81 & \SI{4.9e+02}{} & \SI{-1.6e-13}{} & \SI{2.7e-28}{} & 4.97 & \SI{1.9e+38}{} & \SI{4.2e-08}{} & \SI{1.1e+23}{} & 0.35 & \SI{6.6e+06}{}\\

J1716-0204 & \SI{5.5e-26}{} & -0.77 & 1.48 & 1.97 & \SI{3.0e+02}{} & \SI{-3.1e-12}{} & \SI{1.6e-25}{} & 5.00 & \SI{2.9e+38}{} & \SI{5.1e-07}{} & \SI{1.9e+22}{} & 2.50 & \SI{5.7e+05}{}\\

J1747-0158 & \SI{5.3e-26}{} & -0.11 & 0.99 & 0.65 & \SI{1.1e+02}{} & \SI{-2.5e-14}{} & \SI{2.7e-29}{} & 4.99 & \SI{3.0e+38}{} & \SI{5.3e-07}{} & \SI{3.6e+23}{} & 0.39 & \SI{1.8e+07}{}\\

J1729-0144 & \SI{5.2e-26}{} & 0.13 & 0.47 & 0.77 & \SI{3.3e+02}{} & \SI{-8.8e-13}{} & \SI{1.2e-26}{} & 5.00 & \SI{2.7e+38}{} & \SI{2.2e-07}{} & \SI{4.1e+22}{} & 1.31 & \SI{1.7e+06}{}\\

J1704-0203 & \SI{5.0e-26}{} & 0.71 & 0.88 & 2.96 & \SI{4.5e+02}{} & \SI{-9.9e-12}{} & \SI{1.1e-24}{} & 5.00 & \SI{1.4e+38}{} & \SI{4.7e-07}{} & \SI{1.9e+22}{} & 2.85 & \SI{1.9e+05}{}\\
\bottomrule
\end{tabularx}
\caption{Continuous wave parameters and physical properties of the ten simulated neutron stars with the largest continuous wave strain amplitude $h_0$. }
\label{tab:ns}
\end{table*}

\section{Continuous wave parameter estimation}\label{sec:param-est}
We now investigate using continuous waves to infer properties of neutron stars in the simulated population.
Continuous waves may be detected in various ways, such as through a targeted search of known pulsars, or from a blind all-sky search. In this work we assume that a continuous wave signal has already been detected and sufficiently localised to allow a targeted search with maximum sensitivity.\footnote{This continuous wave signal is assumed to be given by the signal models defined in \cref{eqn:ht,eqn:phi}, have positive $\ddot{f}$, and no unaccounted-for frequency modulation or evolution.}

Previous work has investigated the detectability of continuous waves from a population of neutron stars. \citet{wadeContinuousGravitationalWaves2012} compared the continuous wave amplitude with the estimated noise curve of the detectors; a neutron star is considered detected if its strain amplitude is above the noise curve. 
\citeauthor{wadeContinuousGravitationalWaves2012} also derived a theoretical framework to describe the detectability of continuous waves from neutron stars, which holds for young neutron stars aged $\lesssim 10^7$ years. 
\citet{cieslarDetectabilityContinuousGravitational2021} took into account the spatial position of the neutron star relative to the detector, calculated a signal-to-noise $(S/N)$ ratio, and assumed detection only for stars with $S/N > 11.4$ \citep[based on][]{abbottSettingUpperLimits2004}. 
\citet{reedModelingGalacticNeutron2021a} used the results from existing continuous wave searches \citep{abbottAllskySearchContinuous2019,steltnerEinsteinHomeAllsky2021,dergachevResultsFirstAllSky2020,dergachevResultsHighfrequencyAllsky2021} to obtain a function of strain amplitude and frequency, which is then applied to the simulated population and the ATNF catalogue to determine detectability.
\citet{pitkinProspectsObservingContinuous2011} used Markov Chain Monte Carlo (MCMC) methods to obtain the posterior distributions of signal amplitude parameters, and constraints on the possible range of gravitational quadrupole moment $Q_{22}$ and magnetic field strength $B$, assuming LIGO, Virgo, and Einstein Telescope data.
We simulate continuous wave signals from the synthesised population and use Bayesian inference to estimate the continuous wave signal parameters. From the posteriors of the parameters we then infer the physical properties of the neutron stars and estimate the errors from the inference.

\subsection{Bayesian inference of simulated signals}
From the simulated neutron star population, we select the 10 neutron stars with the largest $h_0$ for the parameter estimation study, which are summarised in \cref{tab:ns}. We note that the first three neutron stars in our population have $h_0$ that exceeds the best upper limit of $h_0\sim \SI{1.1e-25}{}$ ruled out by existing targeted searches using O3 data \citep{abbottAllskySearchContinuous2022}. This could happen as we did not incorporate such restrictions during the simulation. However, the continuous wave frequency range in which the best upper limit is obtained is $f= 100-\SI{200}{Hz}$. None of the top three stars have frequencies in this range, so there is no strong contradiction with the result of existing searches.

We utilise the CWInPy \citep[CW Inference in Python,][]{pitkinCWInPy} package and perform software injections to simulate continuous wave signals. 
Currently, the only third generation detector supported by CWInPy is the D configuration of the Einstein Telescope (ET-D). Hence, we limit the scope of this work to LIGO and ET-D. 
We assume Gaussian noise with a standard deviation defined by the amplitude spectral density specific to each detector at the continuous wave signal frequency $f = 2\nu$. Comparing the sensitivities of LIGO \citep{G1500622-x0,G1401390-v8} and ET-D \citep{Hild_2011,P1600143-v18}, the latter will have noise amplitude spectral density $\sim10$ times smaller than that of LIGO, resulting in higher signal-to-noise ratios.

For LIGO, we simulate detector data at times identical to the real data recorded by each of the two LIGO detectors (Hanford and Livingston) across all observation runs, O1--O3 \citep{LIGOVirg2021-ODFrScObsRAdvLAdV, LIGOEtAl2023-OpDThrObsRLVrKAG}. For the Einstein Telescope, we perform two separate analyses with gap-less detector data spanning 2 years and 5 years respectively. The 2-year analysis is approximately the combined effective observation time of LIGO from O1 to O3b, while the 5-year analysis is similar to the total amount of time spanned by these observation runs (2015-2020), during which the continuous wave signal can evolve. 

The amplitude parameters $\vecA$ and frequency parameters $\vecl$ of the simulated neutron stars are used to generate continuous wave signals in CWInPy. Each simulated signal is then heterodyned and down-sampled to a frequency of $1 /\SI{60}{Hz}$ \citep{2005PhRvD..72j2002D}. 
We perform parameter estimation of both amplitude and frequency parameters, namely $h_0, \cos\iota,\phi_0,\psi,f,\dot{f},\ddot{f}$. 
To obtain the posterior of a parameter, one requires the likelihood, the evidence, and the prior.
The likelihood comes from the Gaussian noise assumption, and we use nested sampling (\citealt{skillingNestedSampling2004, Skil2006-NsSmpGnByCmp}, see \citealt{ashtonNestedSamplingPhysical2022} for an intuitive illustration) to evaluate the evidence and obtain the posterior. The nested sampling scheme is implemented in CWInPy via the Bayesian Inference Library \citep[BILBY,][]{AshtEtAl2019-BUsrByInLGrvAs} which uses \texttt{dynesty} \citep{Speagle-dynesty, Higson2019} for nested sampling.

The choice of priors represents the initial assumptions made for the parameters, which in turn affects the performance of the sampler and consequently the estimated posteriors. 
The prior on $h_0$ is chosen to be a uniform distribution $[ 0,\SI{e -24}{} ]$ to include any likely continuous wave strain amplitude; in the simulated population, $h_0 \lesssim \SI{2e -25}{}$.
The value of $\cos\iota$ is unknown \textit{a priori} and is thus chosen uniformly over $[-1, 1]$. 
The prior on $\phi_0$ is chosen only between $[0, \pi]$, as in CWInPy $\phi_0$ is the initial \emph{rotational} phase offset, and will cover the full phase range when converted to a gravitational wave frequency via $f = 2\nu$. 
The prior on $\psi$ is set to $[0, \pi/2]$ to account for the degeneracy in $\psi$ given by the transformation of $\psi \to \psi+\pi/2$ \citep{jonesParameterChoicesRanges2015}.

For targeted searches, the phase parameters $f=2\nu,\dot{f},\ddot{f}$ are assumed to be roughly determined based on previous observations, e.g.\ from electromagnetic detection of a known pulsar, or from follow-up of a continuous wave candidate found in an all-sky search. 
To ensure a physical braking index, we substitute $\ddot{f}$ for $n$ using \cref{eqn:braking} and impose the constraint $3\le n\le 5$.
We use uniform prior distributions on $f,\dot{f},n$ centred at the true values with single-side widths $3\sigma$, where $\sigma$ is obtained from the inverse Fisher information matrix \citep{jaranowskiSearchingGravitationalWaves2010,luInferringNeutronStar2023}. The priors are chosen to be broad enough to not restrict the posteriors produced by the Bayesian sampler, but narrow enough to allow convergence of the posteriors. Explicitly, $\sigma$ for each parameter is given by:
\begin{align}
    \sigma^2_{f}&= \frac{\mathcal{D}^2}{\pi^2} \frac{1875}{16T^3}, \\
    \sigma^2_{\dot{f}}&= \frac{\mathcal{D}^2}{\pi^2} \frac{1125}{T^5}, \\
    \sigma^2_{n}&= \frac{375 \mathcal{D}^2}{\pi^2} \left( \frac{5 n^2}{16 f^2 T^3} + \frac{3 n (4 n-7)}{\dot{f}^2 T^5} + \frac{420 f^2}{\dot{f}^4 T^7} \right). \label{eqn:sigma-n}
\end{align}
Here, 
\begin{align}
    \mathcal{D}&= \frac{\sqrt{S_h(f)} }{h_0}
\end{align}
is the sensitivity depth, which is related to the signal-to-noise ratio $\rho^2$ by \citep{behnkePostprocessingMethodsUsed2015,dreissigackerFastAccurateSensitivity2018}
\begin{align}
    \rho^2=\frac{4}{25} \frac{h_0^2 T}{S_h}\equiv \frac{4}{25}\frac{T}{\mathcal{D}^2};
\end{align}
$T$ is the observation time; and $S_h$ is the power spectral density of the strain noise in the detector. 

\begin{figure}
    \centering
    \includegraphics[width=\columnwidth]{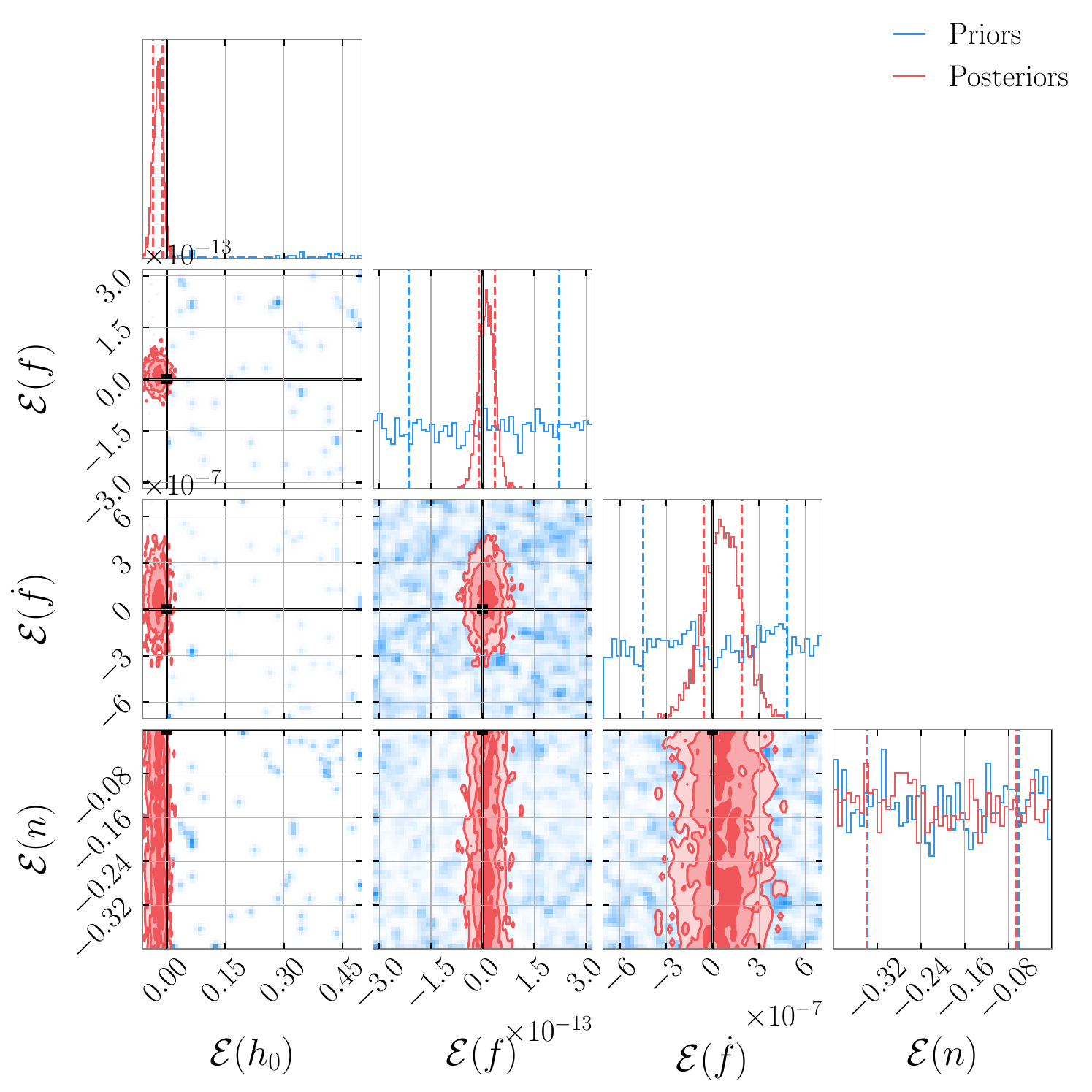}
    \caption{Corner plot of the prior (blue) and posterior (red) distributions of parameter estimation results for an example simulated neutron star (labelled J1704-0203). The ET-D detector with observation time of $T=\SI{2}{yrs}$ is used. Relative errors $\mathcal{E}$ of the posteriors are shown, with true values at $\mathcal{E}=0$. The vertical dashed lines in the 1D histograms represent 1-$\sigma$ confidence, and the three contours in the 2D histograms represent 3-$\sigma$. Note that the distributions of $\mathcal{E}(h_0)$ is zoomed in to make the posterior distribution visible. } 
    \label{fig:PE-corner}
\end{figure}

\Cref{fig:PE-corner} shows a corner plot of the prior and posterior distributions of $h_0,f,\dot{f},n$ for continuous waves from a simulated neutron star in ET-D data with an observation time of 2 years. The 1D distributions for each parameter are shown along the diagonal; the off-diagonal plots show the 2D distributions over pairs of parameters. We plot relative errors $\mathcal{E}$ defined by 
\begin{align}\label{eqn:relerr}
    \mathcal{E}(A)=\frac{A-A^\text{truth}}{A^\text{truth}},
\end{align}
i.e.\ the true values of each parameter are located at $\mathcal{E}=0$. The convergence of the posteriors is consistent with our prior assumption of the detectability of continuous waves, in particular by the Einstein Telescope.
\Cref{fig:PE-corner} also highlights the difficulty in accurately estimating $n$; we will see in \cref{sec:inference} that this limitation dominates the uncertainty in inferring the stellar properties.

\subsection{Inferring physical properties}\label{sec:inference}
Starting with \cref{eqn:Nudot}, \citet{luInferringNeutronStar2023} developed a theoretical framework where the continuous wave signal parameters $h_0,f,\dot{f},\ddot{f}$ and the distance to the neutron star $r$ can be used to infer the physical properties of the star: the principal moment of inertia $I_{zz}$, the ellipticity $\epsilon$, and the perpendicular magnetic dipole moment $m_p$:
\begin{align}
    I_{zz} &=\frac{K_{\mathrm{GW}} c^8 r^2 h_0^2 f}{8 \pi^4 G^2 \dot{f}(3-n)}, \label{eqn:Izz}\\
    \epsilon &=\frac{2 \pi^2 G \dot{f}(3-n)}{K_{\mathrm{GW}} c^4 r h_0 f^3}, \label{eqn:epsilon}\\
    m_p &=\frac{c^4 r h_0}{4 \mu_0\pi^2 G f} \sqrt{\frac{K_{\mathrm{GW}}(n-5)}{K_{\mathrm{EM}}(3-n)}}, \label{eqn:mp} 
\end{align}
where $K_\text{EM}= {2\pi^2}/ {3c^3\mu_0}$ and $K_\text{GW}= {32G\pi^4}/ {5c^5}$.
We assume a $20\%$ error in the measurement of $r$ (i.e.\ $r\sim r^\text{truth}(1+0.2Z)$ where $Z$ is drawn from the standard normal distribution) which is expected to be achievable by the upcoming radio telescope Square Kilometre Array \citep{Smits_2011}.

\begin{figure}
    \centering
    \begin{subfigure}{0.75\columnwidth}
        \centering
        \includegraphics[width=\columnwidth]{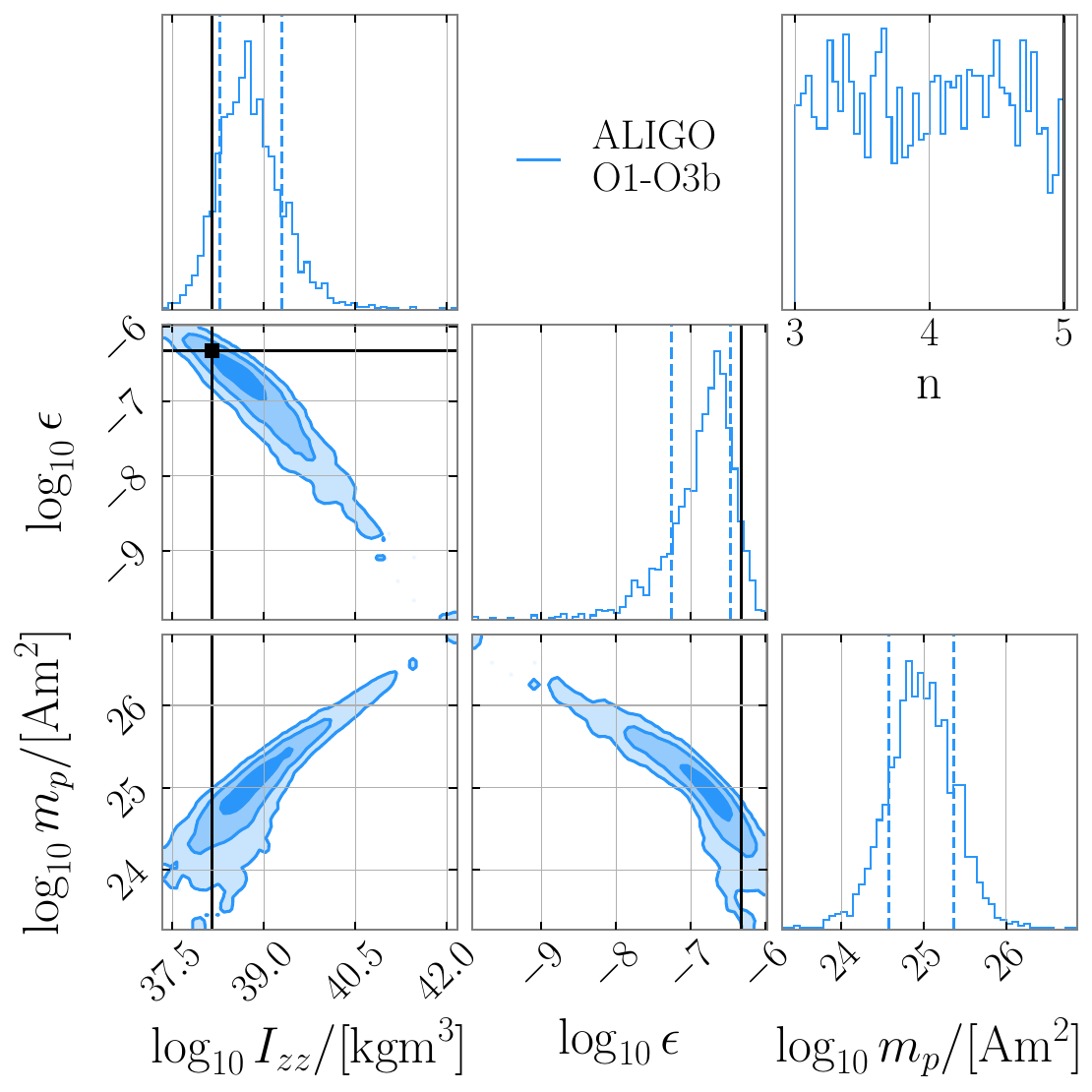}
        \caption{}
        \label{fig:H1_inference}
    \end{subfigure}
    \begin{subfigure}{0.75\columnwidth}
        \centering
        \includegraphics[width=\columnwidth]{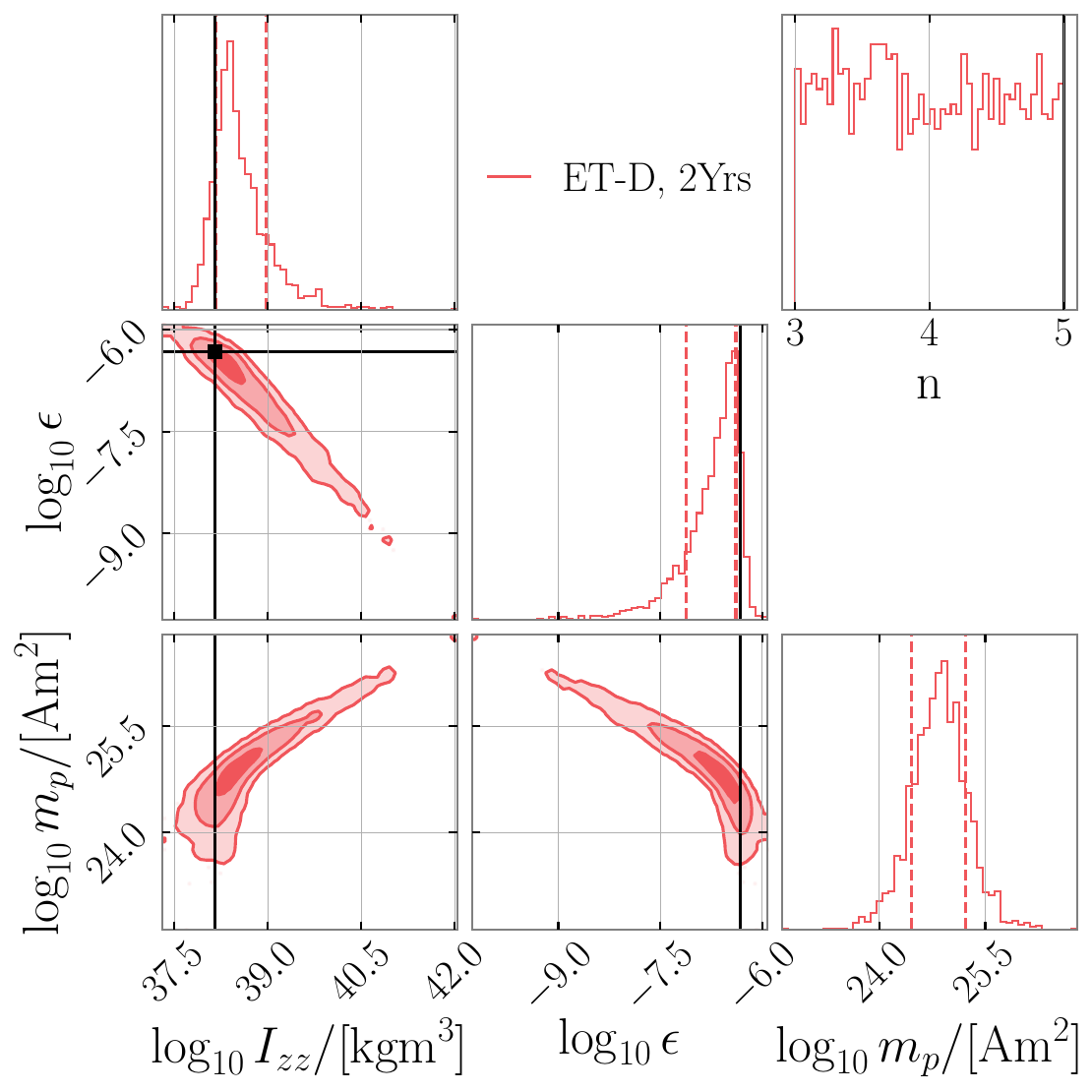}
        \caption{}
        \label{fig:E3_2Y}
    \end{subfigure}
    \begin{subfigure}{0.75\columnwidth}
        \centering
        \includegraphics[width=\columnwidth]{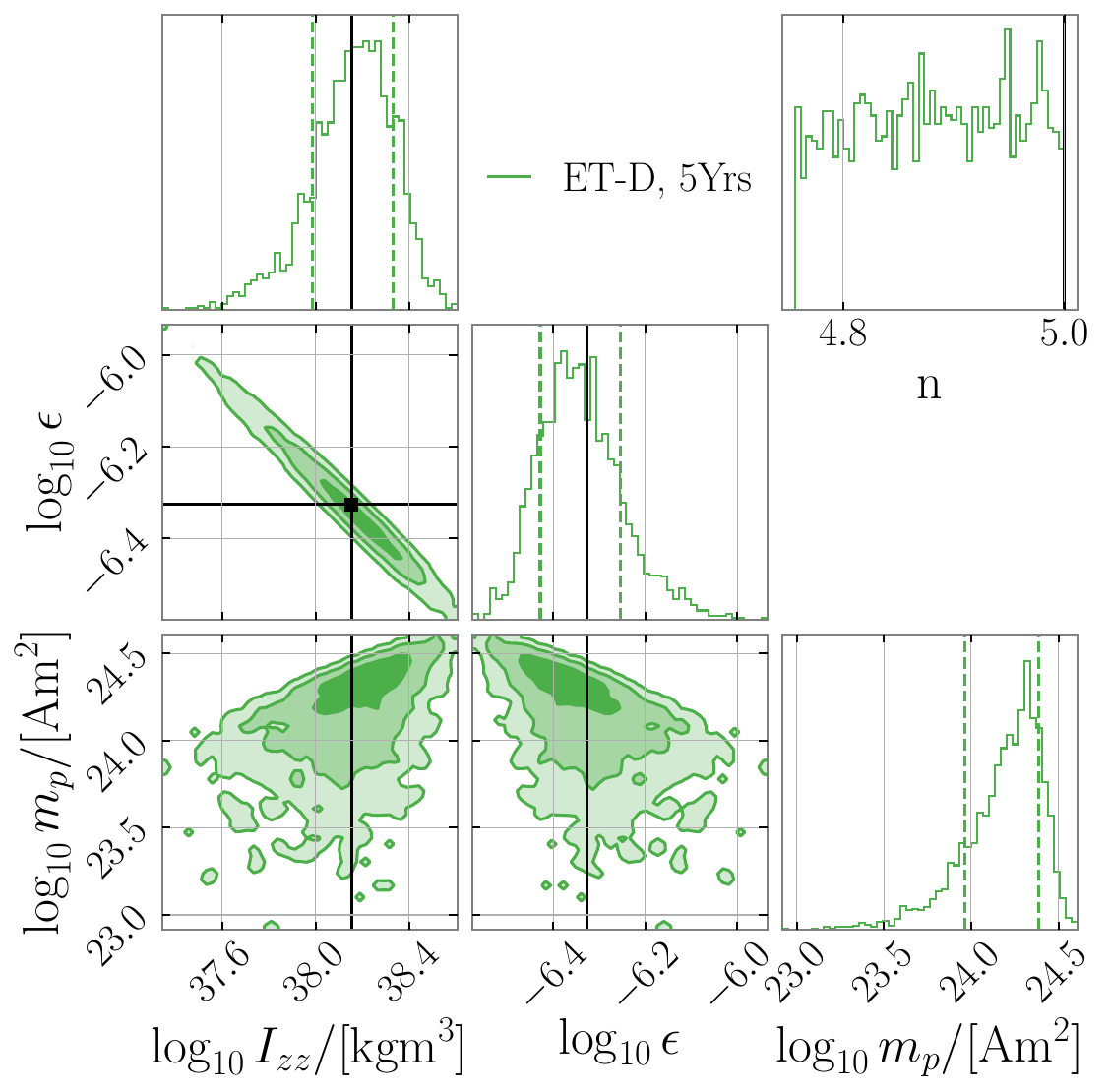}
        \caption{}
        \label{fig:E3_5Y}
    \end{subfigure}
    \caption{Corner plots for the distributions of the logarithm of the principal moment of inertia $I_{zz}$, ellipticity $\epsilon$, and perpendicular magnetic dipole moment $m_p$. Results are for the example simulated neutron star J1704-0203, and (a) Advanced LIGO for the duration of O1-O3, (b) ET-D with $T=\SI{2}{yrs}$, and (c) ET-D with $T=\SI{5}{yrs}$. }
    \label{fig:Inference}
\end{figure}

\Cref{fig:Inference} shows example corner plots of the inferred physical properties, with the posterior distribution of the braking index $n$ positioned at the top right. Comparing \cref{fig:H1_inference} and \cref{fig:E3_2Y}, there is little difference between the posteriors inferred using LIGO versus ET-D with $T=\SI{2}{yrs}$ of data. As seen in \cref{fig:PE-corner}, $f$ and $\dot{f}$ are very well constrained, whereas $\ddot{f}\propto n$ is not: the posteriors of $n$ are uniform over the range $3-5$. (In the 1D histogram of $n$, $n^\text{truth}=5$, and $\mathcal{E}$ ranges from $\mathcal{E}(n=3)=-0.4$ to $\mathcal{E}(n=5)=0.0$). 
This means that the majority of the uncertainty in the inferred parameters stems from the uncertainty in $\ddot{f}$. Since $n$ in both \cref{fig:H1_inference,fig:E3_2Y} is badly estimated, the uncertainties in $I_{zz},\epsilon,m_p$ are similar for both LIGO and ET-D at $T=\SI{2}{yrs}$.
With an observation time of $T=\SI{5}{yrs}$, however (\cref{fig:E3_5Y}), $\sigma_n$ [\cref{eqn:sigma-n}] is small enough so that the prior on $n$ is a narrow subset of $[3,5]$, and consequently the uncertainty on $I_{zz}$ and $\epsilon$ is improved.

\begin{figure}
    \centering
    \includegraphics[width=0.8\columnwidth]{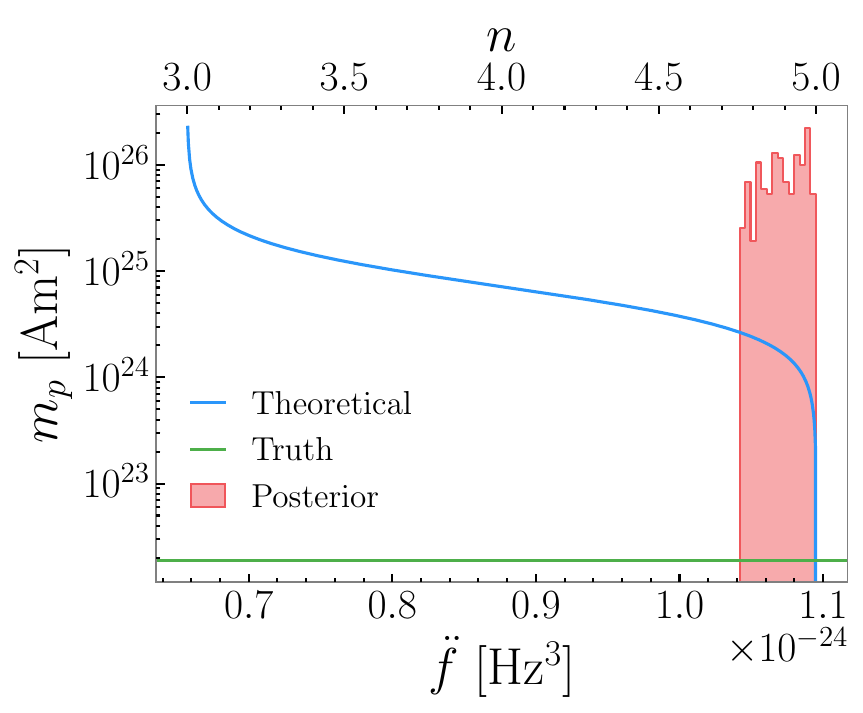}
    \caption{Theoretical model by \citet{luInferringNeutronStar2023} of $m_p$ as a function of $\ddot{f}$ on top of the posterior distribution of $n$ for ET-D at $T=\SI{5}{yrs}$. The true value is plotted in green. }
    \label{fig:BehavioursMp}
\end{figure}

The inference of $m_p$ is challenging due to the behaviour of \cref{eqn:mp}. \Cref{fig:BehavioursMp} plots the theoretical model for $m_p$ as a function of $\ddot{f}$ (or equivalently $n$). The posterior of $n$ is shown in red, and the true $m_p$ is plotted in green. Note the asymptotic behaviours of $m_p$ at $n=3$ and $5$; consequently, when a neutron star has $n\approx 5$, any small deviation of the estimated $n$ will drastically affect the estimation of $m_p$. Therefore, while the posterior of $m_p$ is supposed to contain the true value since the posterior of $n$ contains the true value, in reality, the fact that we obtain the posterior through sampling, coupled with the asymptotic behaviour of $m_p$, makes it unlikely that any sampled $n$ will be close enough to $n^\text{truth}$ to produce a good estimate of $m_p$.
This is unsurprising: given that we are inferring neutron star properties from continuous wave detections, we expect most of the detected neutron stars to have a continuous wave-dominated spindown with $n \approx 5$. 
As the magnetic dipole moment is related to the strength of the electromagnetic radiation, which is not the dominant form of radiation for these stars, the amount of energy emitted via electromagnetic radiation, and hence the ability to infer $m_p$ through this method, is limited for neutron stars with $n \approx 5$. 
Similarly, the skewed distributions of $I_{zz}$ and $\epsilon$ in \cref{fig:H1_inference,fig:E3_2Y} are also a consequence of the theoretical framework [\cref{eqn:Izz,eqn:epsilon}] and the posterior on $n$. Both equations have asymptotes at $n=3$, with $\lim_{n\to 3^+} I_{zz}=+\infty$ and $\lim_{n\to 3^+} \epsilon=0$. 
As the posterior on $n$ approaches 3, therefore, $I_{zz}$ and $\epsilon$ will increasingly diverge from their true values. Unlike $m_p$, however, a much higher fraction of posterior samples (when $n > 3$) produce good estimates of $I_{zz}$ and $\epsilon$; as seen in \cref{fig:Inference}, the highest densities in the $I_{zz}$--$\epsilon$ distributions coincide with the true values for both LIGO and ET-D.

\begin{figure}
    \centering
    \begin{subfigure}{0.9\columnwidth}
        \centering
        \includegraphics[width=\columnwidth]{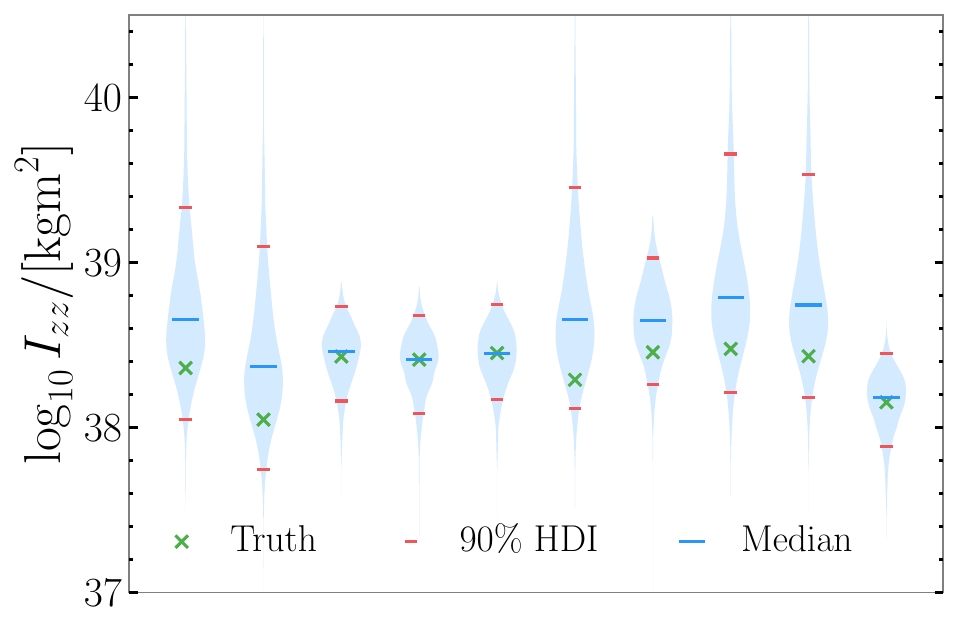}
        \caption{}
        \label{fig:ViolinIzz-pdf}
    \end{subfigure}
    \begin{subfigure}{0.9\columnwidth}
        \centering
        \includegraphics[width=\columnwidth]{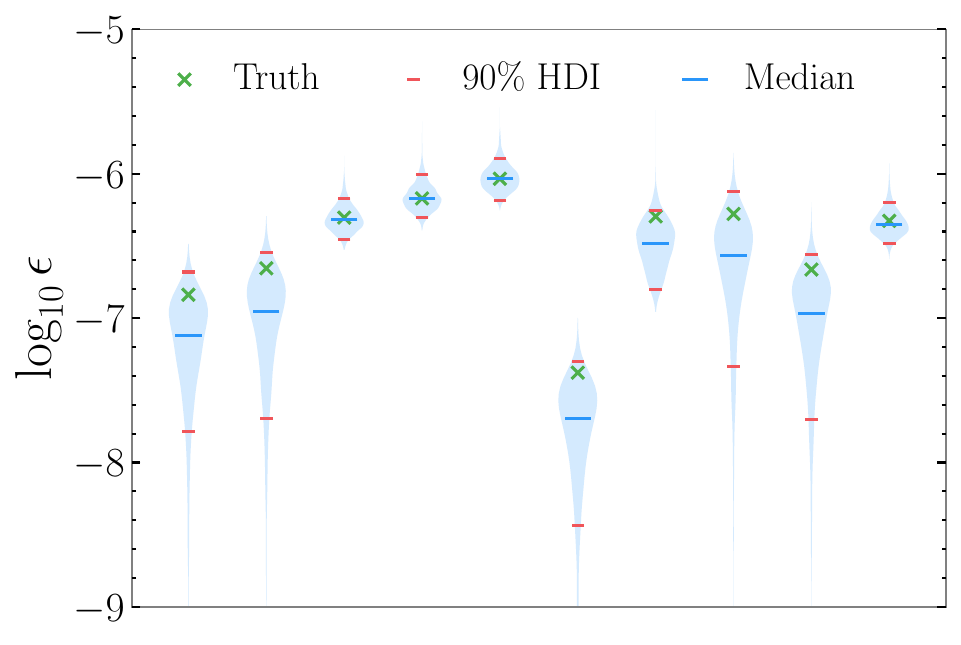}
        \caption{}
        \label{fig:ViolinEllip-pdf}
    \end{subfigure}
    \begin{subfigure}{0.9\columnwidth}
        \centering
        \includegraphics[width=\columnwidth]{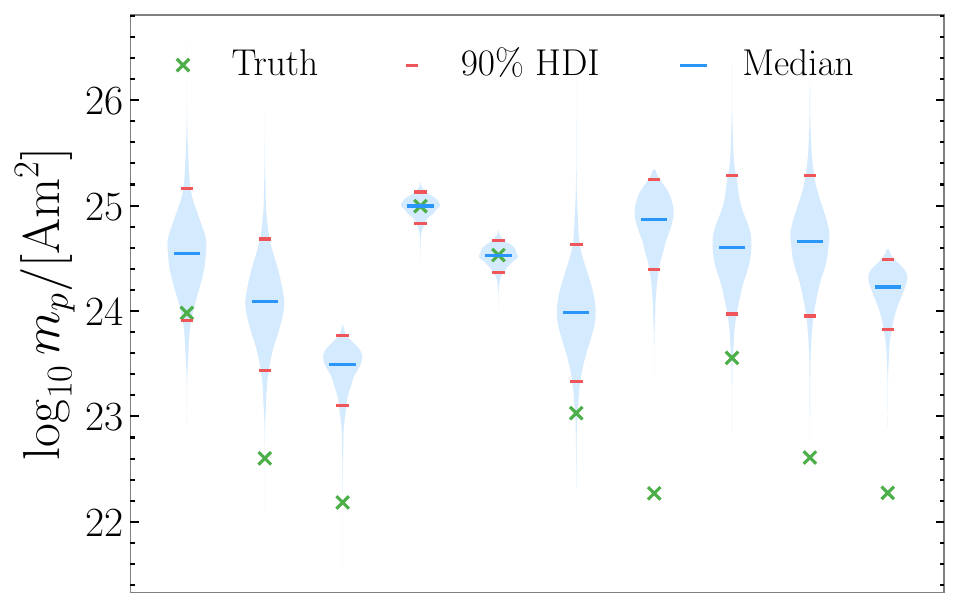}
        \caption{}
        \label{fig:ViolinMp-pdf}
    \end{subfigure}
    \caption{Violin plots showing the distributions of (a) $I_{zz}$ , (b) $\epsilon$, and (c) $m_p$ for the parameter estimations with CWInPy on ten neutron stars, arranged in descending order of $h_0$. The $90\%$ highest density intervals from each run are shown in red; the medians are shown in blue; and true values are shown as green crosses. ET-D is used with $T = \SI{5}{yrs}$.}
    \label{fig:Violins}
\end{figure}

\begin{figure}
    \centering
    \includegraphics[width=\columnwidth]{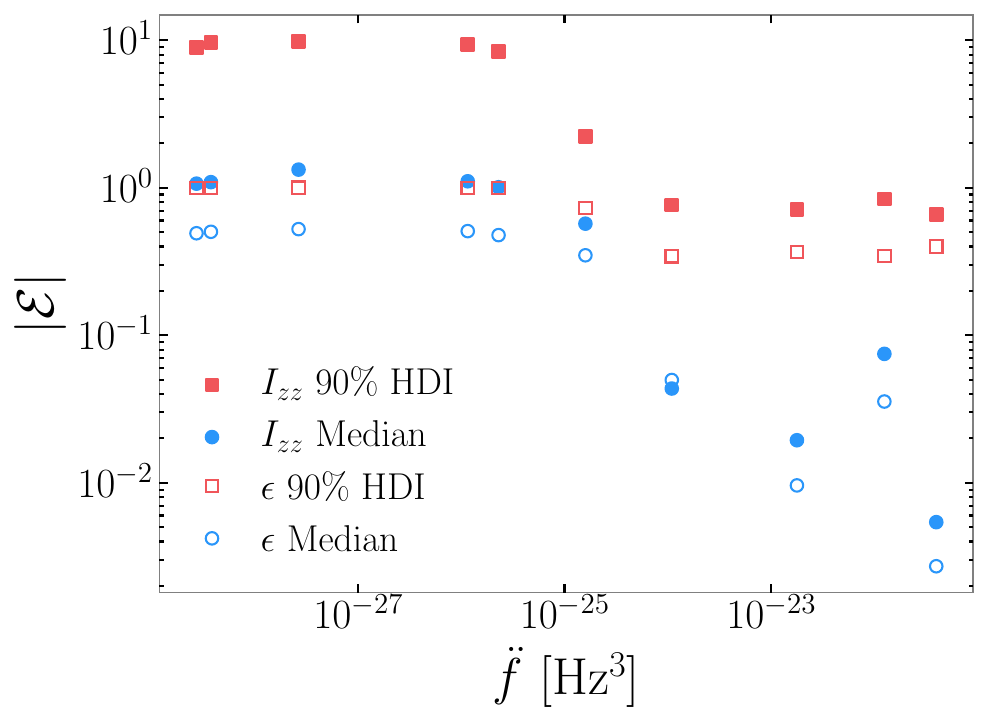}
    \caption{Magnitudes of relative errors in $I_{zz}$ (filled markers) and $\epsilon$ (hollow markers) plotted against $\ddot{f}$ for ET-D with $T = \SI{5}{yrs}$. The maximum errors of the 90\% HDI are shown in red, and that of medians are shown in blue. }
    \label{fig:relerr_trend}
\end{figure}

In \cref{fig:Violins}, we show the inferred physical properties of the 10 neutron stars with the largest $h_0$ in the simulated population, in descending order of $h_0$. The posterior distributions are shown in the form of violin plots, in which the widths of the ``violins'' give the posterior probability densities. The red lines represent the \emph{highest density intervals} (HDI) of the distributions, defined as the smallest possible $90\%$ credible intervals, i.e.\ the smallest bounds we can place on the physical properties. 
The varying sizes of the distributions are due to the different extents to which $n$ is constrained. For neutron stars where $n$ is well constrained, the uncertainties in the inferred properties are also smaller. 
This is confirmed in \cref{fig:relerr_trend}: for neutron stars with higher $\ddot{f}$, the uncertainties in the inferred parameters, as quantified by the relative error $\mathcal{E}$, are smaller. This can be understood by considering the continuous wave signal searched for by CWInPy. The contribution of $\ddot{f}$ to the frequency evolution of a continuous wave is more pronounced if $\ddot{f}$ is larger, and hence CWInPy can more readily constrain larger $\ddot{f}$. 

From \cref{fig:relerr_trend} we see that, for five years of continuous wave observations using ET-D, the physical properties of the 10 neutron stars in the synthesised population have errors that depend on the estimation of $n$, with $\ddot{f}$ being a contributing factor. For neutron stars with small $\ddot{f}$, $I_{zz}$ can be inferred with an error of $\sim100\%$ for point estimates using the median. The 90-th percentile credible interval yields a maximum error of $\sim1000\%$. For the ellipticity $\epsilon$, a point estimate with relative error $\sim50\%$ can be made, and the 90-th percentile credible interval giving a maximum error of $\sim100\%$. 
For neutron stars with more significant $\ddot{f}$, the median of the posterior of $I_{zz}$ is accurate enough to be within $\sim10\%$ of the true value, while the 90-th percentile credible interval gives a maximum error of $\sim100\%$. The ellipticity $\epsilon$ can be estimated using median with a relative error of $\sim 5\%$, and a maximum error from the credible interval of $\sim 50\%$.
Few alternative methods exist to measure $I_{zz}$. Separate measurement of neutron star mass and radius can provide a measurement of $I_{zz}$, but measuring these two properties simultaneously is challenging \citep{steinerUsingNeutronStar2015,millerPSRJ003004512019}. Another technique measures $I_{zz}$ through the higher-order relativistic correction to the periastron advance of rapidly spinning binary pulsars \citep{damourHigherorderRelativisticPeriastron1988}. To date, this approach can only be applied to the double pulsar PSR J0737-3039, with the measurement yielding errors of $\sim 10$--20\% \citep{miaoMomentInertiaPSR2022}, comparable with our point estimates  of $\sim 10\%$ for neutron stars with large $\ddot{f}$. 
So far, no alternative approach to measuring $\epsilon$ is available other than through a continuous wave detection \citep{luInferringNeutronStar2023}. We are unable to accurately infer the perpendicular magnetic dipole moment, but it may be inferred through alternative methods, such as from pulsars directly through the measurement of frequency and spindown \citep{kramerPulsars2005}.

\section{Discussion}\label{sec:discussion}
This paper presents an analysis of the capability of LIGO and ET to measure physical properties of neutron stars using continuous gravitational waves. We first synthesised a population of neutron stars using Monte-Carlo techniques, performed parameter estimation using Bayesian inference on the ten gravitationally loudest simulated continuous wave signals, and finally inferred their physical properties, namely the stellar moment of inertia $I_{zz}$, equatorial ellipticity $\epsilon$, and the perpendicular magnetic dipole moment $m_p$.

Targeted searches for continuous waves from the synthesised neutron stars produced well-constrained posteriors for the continuous wave strain amplitude $h_0$, frequency $f$, and its first order time derivative $\dot{f}$, for both LIGO and ET-D. The inference of the braking index $n$ proved challenging, and was the cause of the majority of the uncertainty in the inference of the physical properties $I_{zz},\epsilon,m_p$, but this can be improved with longer observation periods. 
Using the ET-D configuration with five years of observations, depending on estimation of $n$, which is related to the size of $\ddot{f}$, 90-th percentile credible intervals can be placed on $I_{zz}$ and $\epsilon$ with errors of $\sim100-1000\%$ and $\sim50-100\%$, respectively; and point estimates using median can be made with errors of $\sim10-100\%$ and $\sim5-50\%$, respectively. The perpendicular magnetic dipole moment could not be properly inferred for neutron stars with $n\approx 5$ due to the asymptotic behaviour of \cref{eqn:mp}.

\citet{luInferringNeutronStar2023} found that measurement of $I_{zz}$, $\epsilon$, and $m_p$, with relative errors of $\lesssim 27$\%, might be achievable with continuous waves.
In comparison, our headline results (summarised above) are somewhat less promising for the accurate estimation of these parameters.
The reason for this discrepancy is readily apparent from \cref{tab:ns}; the 10 neutron stars used for parameter estimation all have relatively small spindowns ($|\dot{f}| \lesssim \SI{1e-10}{Hz/s}$, $|\ddot{f}| \lesssim \SI{1e-22}{Hz/s^2}$) compared to what is typically assumed for continuous wave searches. Indeed, as seen in Fig.~4 of \citet{luInferringNeutronStar2023}, the $|\dot{f}|$ of the 10 neutron stars are similar to (or even smaller than) the smallest $|\dot{f}| \sim \SI{1e-11}{Hz/s}$ covered by \citet{luInferringNeutronStar2023}, where errors were found to be worst ($\mathcal{E} \gtrsim 30$). Our results, therefore, reaffirm the intuition that the detection, not only of continuous waves, but from a relatively luminous neutron star with high spindown, would be required for any useful estimation of its physical parameters.

Future work could improve upon some of the simplifying assumptions used in the neutron star population synthesis, such as performing $n$-body simulations of the spatial evolution of the neutron stars through the Galactic potential, or adopting evolution models for the ellipticity and the magnetic field strength. In addition, directed or all-sky searches could be performed on the synthesised population to provide a more comprehensive assessment of the detection prospects for LIGO and the Einstein Telescope.

\section*{Acknowledgements}
We thank Julian Carlin for helpful comments on the manuscript.
Y.H., K.W., and S.M.S.\ are supported by the Australian Research Council
Centre of Excellence for Gravitational Wave Discovery (OzGrav) through project
number CE170100004.
During part of this work M.D.P was supported through the UK Science \& Technology Facilities Council grant ST/V005707/1.
The parameter estimation study was performed on the OzSTAR national facility at Swinburne University of Technology. The OzSTAR program receives funding in part from the Astronomy National Collaborative Research Infrastructure Strategy (NCRIS) allocation provided by the Australian Government, and from the Victorian Higher Education State Investment Fund (VHESIF) provided by the Victorian Government.
This material is based in part upon work supported by the LIGO Laboratory which is a major facility fully funded by the National Science Foundation.
This manuscript has document number LIGO-P2300244.

\section*{Data Availability}

The data underlying this article will be shared on reasonable request to the
corresponding author(s).

\bibliographystyle{mnras}
\bibliography{bib}
\bsp	% typesetting comment
\label{lastpage}
\end{document}